\documentclass[12pt,preprint]{aastex}

\slugcomment{}

\shorttitle{Risaliti et al.}
\shortauthors{Variability of N$_H$ in Seyfert Galaxies}

\begin{document}

\title{Ubiquitous variability of X-ray absorbing column densities in Seyfert 2 
Galaxies}

\author{G. Risaliti\altaffilmark{1,2,3}, M. Elvis\altaffilmark{1}, F.
Nicastro\altaffilmark{1}}  

\email{grisaliti@cfa.harvard.edu}

\altaffiltext{1} {Harvard-Smithsonian Center for Astrophysics,
    60 Garden Street, Cambridge, MA 02138} 
\altaffiltext{2} {Dipartimento di Astronomia, Universit\`a di Firenze, L.go
E. Fermi, 5, I-50125 Firenze, Italy}
\altaffiltext{2} {Osservatorio Astrofisico di Arcetri, L.go
E. Fermi, 5, I-50125 Firenze, Italy}     

\begin{abstract}
We present a study of the variations in the absorbing column density of
25 X-ray defined Seyfert 2 galaxies, as inferred from hard X-ray observations, on timescales
from months to several years. We show that a significant variation of N$_H$
(from 20\% to 80\%) is observed in almost all (22/25) the sources with multiple
X-ray observations, although X-ray absorption never vanishes. For
a subsample of 11 sources observed
at least five times the typical variation time, as defined by a structure function,
is less than one year for both heavily absorbed (N$_H \sim 10^{23}$ cm$^{-2}$)
and moderately absorbed (N$_H \sim 10^{22}$ cm$^{-2}$) sources.
These variations rule out the simplest version of the unified models, based
on a homogeneous obscuring torus, and suggest the presence of
clumpy
circumnuclear material on a scale well below a parsec. We propose a modification
of the torus model in which an overabundance of slightly dusty BELR clouds obscures
the BELR. The BELR needs, like the torus, to have an axisymmetric structure.
This model is closely related to that of Elvis (2000) for type 1 AGN.
For lightly obscured AGN (N$_H \sim 10^{22}$ cm$^{-2}$) the structure function
shows an increase at a timescale of $\sim$5 yr, indicating a second absorber, most
probably on a 5-10 pc scale associated with the host galaxy.
\end{abstract}


\keywords{}

\section{Introduction}

Much observational evidence suggests that strong obscuration absorbs the strong continuum
source in Active Galactic Nuclei
over a significant
fraction of the solid angle (Lawrence \& Elvis 1982, Antonucci \& Miller 1985, Maiolino et al. 1998).
The effects of this absorption are clearly visible in the
X-ray spectra of many AGNs, where a   photoelectric cutoff  at energies of a keV or so
(depending on
the column density of the gas) is observed, and in the mid-infrared, where we observe the
thermal re-emission of the absorbed UV radiation from the dust associated with the
absorbing gas.This circumnuclear material obscures the optical emission
of the Broad Line Region, but not that of the more distant Narrow Line Region. As a
consequence, objects observed along a line of sight covered by this material appear
in the optical as type 2 AGNs.

The simplest geometry for this cold gas surrounding the nucleus,
is that of a torus covering $\sim 80$\% of the solid angle (Antonucci \& Miller 1985, Krolik \& Begelman 1989).
Axial symmetry in the material
in the central $\sim$ 100 pc of nearby AGNs is suggested by the biconical shapes (Pogge 1989) seen well in
high resolution
HST images (Tadhunter \& Tsvetanov 1989 , Malkan et al. 1998), and
polarization requires a highly non-spherical shape (Antonucci \& Miller 1985).
One of the unsolved questions about the
structure of this putative torus is its typical dimensions. 
Detailed models of homogeneous circumnuclear tori,
both on the 1 pc and 100 pc scale, reproduce the observed infrared
SED (Pier \& Krolik 1992, Granato \& Danese 1994).
Hard X-ray observations
show that at least 50\% of nearby Seyfert 2s are obscured by a column density
higher than 10$^{24}$ cm$^{-2}$ (Risaliti et al. 1999). In these cases it is 
unlikely that the typical dimensions of a homogeneous torus exceed a few parsecs, since otherwise
the dynamical mass of the obscuring material would be too large (Risaliti et al. 1999).

However, detailed observations both in the X-rays and in the near IR/optical band suggest
that the structure of the circumnuclear material is more complex than a simple homogeneous
torus. Large variations in the cold obscuring column density have been known
in a few objects for many years (Ives et al. 1976, Warwick et al. 1988).
A more systematic study by Malizia et al. (1997),
shows that variations of several 10$^{21}$ cm$^{-2}$ are rather common in both type 1
and type 2 Seyferts (17 out of 23 sources observed at least 3 times shows N$_H$ variations).
Changes in months or less suggest much smaller sizes, whether the observed
change is due to motions across the line of sight, or to a varying ionization state. The broad emission
line region (BELR) has been suggested as a site for this variable absorption in type 1 AGNs (Ives et al. 1976).
Since the BELR
is either hidden or not present in type 2 AGN this explanation appears implausible for type 2
objects, which comprise 80\% of all AGN (Maiolino \& Rieke 1995).
In this paper we investigate the variability of the X-ray absorbing column density in 
X-ray-defined
Seyfert 2s having column densities higher than $\sim 10^{22}$ cm$^{-2}$, but less than 10$^{24}$ cm$^{-2}$,
i.e. Compton thin. This includes optically defined
both types 1.8 and 1.9 objects (which show broad lines only in long wavelength lines, e.g. H$\alpha$,
Pa$\alpha$, Osterbrock 1989) as well as type 2 Seyferts, and all show evidence for cool
absorbing material.

We collected all the data available in the literature for Seyfert 2s and we complemented
them with the analysis of unpublished data in the ASCA and BeppoSAX public archives.
We found that a sample of 25 sources were observed at least twice in the hard X-rays.
11 objects out of these 25 have 5 or more hard X-ray observations in a time interval of
several years, and we can therefore
obtain for them an ``N$_H$ light curve''.

In the next Section we show that variations in N$_H$ are
almost universally present. We analyze in greater detail the well-observed 11 sources, dividing
them in two subsamples with N$_H
\sim 10^{23}$ cm$^{-2}$ (6 objects) and   N$_H \sim 10^{22}$ cm$^{-2}$ (5 objects). Previous
X-ray studies show that sources
with 10$^{22}$ cm$^{-2} <$N$_H < 10^{23}$ cm$^{-2}$ are mostly type 1.8 and 1.9 (Risaliti
et al. 1999) while those with N$_H>10^{23}$ cm$^{-2}$ are mostly pure Seyfert 2s. This
dichotomy is also present in our sources: 3 out of 5 objects in the low-N$_H$ subsample are
intermediate type, while 5 out of 6 in the high-N$_H$ subsample are type 2s (Table 1).

In Section three we analyze the N$_H$ variability of the well studied subsample
using a ``Structure Function'' similar to that used in studies of brightness variability
of quasars (Fiore et al. 1998, Di Clemente et al. 1996).
In Section four we discuss our results and we compare them with current models of the
circumnuclear medium of AGNs. Our conclusions are summarized in Section 5.

\section{Data}

In Table 1 we list all 139 measurements of N$_H$ (and 2-10 keV flux) in the literature (with references) for all
the 25 Seyfert 2s with N$_H > 10^{22}$ cm$^{-2}$ that have been observed at least twice.
Quite a fraction (17 \%) of the
observations have not previously been published, and their inclusion strengthens our results significantly.
Where BeppoSAX or ASCA measurements are not published,
we measured the N$_H$ values through a standard analysis of the data from the
BeppoSAX SDC public archive and the ASCA public archive at HEASARC.
More details of this analysis will be given in
Risaliti (2001, in preparation, see also Appendix A).

For this and subsequent
analysis, we used the relevant data from every past X-ray mission. We note that
this should not give problems of intercalibration, since the column density is
evaluated from the energy of the steep photoelectric cutoff in an X-ray
spectrum. Indeed, while the flux calibration between X-ray instruments is notoriously
uncertain, the calibration in energy is quite good, especially at energies higher
than 1-2 keV, where the instrumental responses do not vary rapidly, and where cut-offs for the N$_H$ range of
these Seyfert 2s are found.

Figure 1 shows the factor {\it f} by which the column density changes in the sample objects. {\it f}
is defined as the ratio between the
measured variation of N$_H$ and the mean of the two values.
When more than two observations
are available, we chose according to both the size of the variation and the statistical
significance: if at least two measurements were different at a confidence level 
higher than 90\%, we chose those  with the largest factor variation. Otherwise, we selected the
measurements with the most statistically significant variation. 

The main result of our analysis is clear from Fig. 1: a significant column density
variation on a timescale of years 
is not a peculiar property of a few sources, but is virtually ubiquitous
in Seyfert 2s: only 3 of the 25 Seyfert 2s are consistent with no variation. These have only
2 or 3 observations in the hard X-rays so that variations typical of well-observed sources
could easily be missed.

In Figure 2 we plot the relative N$_H$ and relative 2-10 intrinsic keV flux for each observation
of the well observed subsample.
Both the measurements are normalized to the average values for each source. The intrinsic flux has been obtained
using the best fit values for each observation, and assuming no absorption. Interestingly, no correlation
between variations of N$_H$ and flux appear.
If photoionization plays an important role in N$_H$ variations, we would expect that
N$_H$ values below the average of a given source (values $<$ 1 in Fig. 4) are associated to
flux measurements above the average ( values $>$ 1), and vice versa. Therefore, the regions
with $f_x > 1$, $f_y <1$ and   $f_x < 1$, $f_y >1$ (top-right and bottom-left quadrants) should
be more populated than the regions  $f_x > 1$, $f_y >1$ and  $f_x < 1$, $f_y <1$ (top-left and bottom-right
quadrants). Instead, the density of points in the four regions is about the same, as expected if
flux and N$_H$ variations are uncorrelated. Moreover, if photoionization effects were important
for an object, we would expect a correlation among the points relative to this object.
Instead, all the 11 groups of points in Fig. 4 (each of which is marked with a different
symbol) appear to be randomly distributed in the N$_H$-flux plane.
As a further confirmation, we quantitatively estimated the effects of photoionization
assuming two different values of the ionization parameter, U=0.1 and U=0.5, at the inner edge
of the absorber.
We assumed that the absorber has a total column density N$_H=10^{23}$ cm$^{-2}$,
distributed on a region $\sim 5$ times thicker that the distance of the inner edge from the
center. These parameters are quite extreme for a ``standard'' absorber on the parsec
scale. They are reasonable if we assume that the absorber is located in the Broad Emission
Line Region (BELR). We believe this is a likely possibility, as we show in the following sections.
In any case, this scenario can be regarded as extreme in the sense that within it photoionization
plays a more important role than in standard torus models.
The result we obtained is that fitting the photoionized gas with a cold absorption model,
the equivalent cold N$_H$ changes only of $\sim 5$\% if the ionization parameter changes from
U=0.1 to U=0.5 (a strong continuum variation). Therefore, we conclude that even strong
variations in the continuum are not able to cause apparent variations of  cold N$_H$ higher than a few
10$^{21}$ cm$^{-2}$. This further rules out photoionization as a cause of the observed
absorption variations.

In order to analyze in greater detail the timescale and the nature of the observed
N$_H$ variations, we focused on a subsample of 11 bright Seyfert 2s with multiple
observations in the hard (2-10 keV) X-ray band. Conveniently  these 11 sources are divided
into two groups: 5 (marked `A' in
Table 1) have N$_H \sim 
10^{22}$ cm$^{-2}$, while the other 6 (marked `B') have N$_H \sim 10^{23}$ cm$^{-2}$.
This division is in rough agreement with the optical classification of type 1.8/1.9 for
low-N$_H$ objects and type 2s for high-N$_H$ objects (Table 1, Risaliti et al. 1999)

The ``light curves'' of N$_H$ are shown for samples A and B in Figures 3 and 4, respectively.
The increasing or decreasing
of N$_H$ is not correlated with the instruments used for the measurements, reinforcing
the conclusion that the effect is
not due to problems in instrumental calibration.
Figures 3 and 4 seem
to show variations on a variety of timescales from months to years. In Table 2 we list the fastest
change in N$_H$ for each object in the full sample, detected at 90\% significance or greater.
The objects are sorted according to the observed variation timescale. We note that the
shortest variation times are observed in the most extensively studied objects: only 3 out of the
11 sources with 5 or more X-ray observations show variations only on timescales of several years.
Therefore, the observed N$_H$ variability timescales in many objects are only a conservative  upper limit:
the fastest variations are not detected only because the time delays between the existing
observations are too long.

\section{Structure Function}

The data for the two well observed subsamples
can be used to obtain a ``column density
structure function'', defined by analogy with the intensity structure function
used to study optical and X-ray variability of QSOs (Di Clemente et al. 1996, Fiore
et al. 1998).
We considered 10 time intervals, $t_k$,
for the low N$_H$ subsample ``A'' and 11 for the high N$_H$ subsample ``B'',
from 1 to 20 years.
We
defined a {\em structure function}, $F(t_k)$, as the average ratio of all the N$_H$
measurements separated by a time between $t_k$ and t$_{k+1}$ for each set of 5 sources:

\begin{equation}
F(t_k) = \frac{1}{N_k} \sum_{(i,j)} {\rm \vspace{1cm}}   \sum_{l(t_k)}
\frac{{\rm max} (N_H(j),N_H(l))}{{\rm min} (N_H(j),N_H(l))} 
\end{equation}

where {\em i} goes from 1 to 5 in subsample A and from 1 to 6 in subsample B, and labels the source; {\em j}
labels the N$_H$ measurements for each
source, while the sum over {\em l} is extended to data N$_H$(l) for which the time delay
from N$_H$(j) is between  $t_k$ and $t_{k+1}$.  $N_k$ is the total number of pairs
of measurements separated by a time between $t_k$ and $t_{k+1}$.
The width of the  time intervals was chosen in order to have at least
15 data points per bin.

Since the ratio in the above equation is always greater than
1, the value is biased positive. 
Following Di Clemente et al. (1996)
we remove this bias by defining a modified structure function as follows:

\begin{equation}
F'(t_k) = \sqrt{ F(t_k)^2 - \frac{2}{\pi}<\sigma_{k,l}>^2}
\end{equation}
where $<\sigma_{k,l}>$ is the average of the statistical errors on the ratios in Eq. 1.

The Function $F'(t_k)$ is plotted in Fig. 5a for the low-N$_H$ subsample (A) and in
Fig. 5b for the high-N$_H$ sample (B). We removed the points of signal-to-noise less than
3$\sigma$
to avoid an unphysical decrease of $F'(t_k)$ due to
noise. We also used only one of the high 5 signal-to-noise  points provided by BeppoSAX observations
of the bright source Centaurus A. These observations show a low N$_H$ variation (although significant
at a level higher than 90\%), but lower than that measured in average in the other
observations. Including all
the 5 observations would bias our results since  the statistics would be
dominated by the single source Centaurus A.

From Fig. 5a and 5b and Fig. 2 we can draw the following conclusions:
\begin{enumerate}
\item The variability is statistically highly significant for both the low-N$_H$ and the
high-N$_H$ objects at all the timescales considered (90 days - 20 years).
\item For both samples the variability is already significant at the shortest
timescale that can be investigated with our data (several months). This result is also
confirmed by the data in Table 2 for to the whole sample:
sources with measured variations in a time shorter than 1 year are
$\sim$ 50\% of the whole sample, rising to $\sim$70\% of the subsample of sources
for which we actually have observations within a time interval of less than 1 year.
\item 
The modified structure function is constant (within
statistical errors) for the high N$_H$ sample, while for the low N$_H$ sources it seems
to have a significant increase for $t_k > $ 5 years. This suggests a characteristic
timescale of $\sim$ 5 years for variations in the order of 10$^{22}$ cm$^{-2}$, since
the structure function is expected to increase with time when a slower variation is
convolved (randomly) with faster ones.
\item
Despite the large N$_H$ variability observed, no object has been observed to change from type
1 to type 2 (i.e. N$_H$ never drops below several 10$^{21}$ cm$^{-2}$)
\item Variations of N$_H$ are not correlated with X-ray flux variations (Fig. 2).
\end{enumerate}

\section{Discussion}
\subsection{The distance from the center of the obscuring gas}
In the previous Section we demonstrated that N$_H$ variability in a timescale from
months to years is present in the great majority, and possibly in all, Seyfert 2 galaxies.
The structure function described in the previous section is useful to
quantify the general N$_H$ variability properties of our sources.
A physical interpretation of our results must explain both the shape
of this function and the different properties of the individual sources.

There are two physical reasons that can explain the variability of the absorbing column density:
variations in the ionization state of the absorber, due to variations in the ionizing radiation,
and variations in the amount of absorbing gas along the line of sight. In the first case, the absorber
can be homogeneous, and the variation in N$_H$ should be correlated with intrinsic flux variations.
In the second case, the absorber must be clumpy, and the variation timescales will be
correlated with the typical crossing time of an absorbing cloud along the line of sight.
However we already showed that a change in ionization does not fit the data (Section 2, Fig. 2)
We can therefore adopt the second scenario -motions in a clumpy medium-
and use the information collected in the previous sections
to estimate the distance from the center of
the obscuring gas.

We can idealize the situation by assuming the typical timescale of variation, {\it t},
to be the crossing time of a discrete cloud across the line of sight.
Assuming that the absorption is due to spherical clouds moving
with Keplerian velocities, the distance from the central black hole of mass {\it M} is given by:
\begin{equation}
R \sim 3 \times 10^{16} \frac{M_\bullet}{10^9M_\odot}~(\frac{\rho}{10^6 {\rm cm}^{-3}})^2~(\frac{t}
{{\rm 5~ Msec}})^2~(\frac{N_H}{10^{22}{\rm cm}^{-2}})^{-2} ~{\rm cm}
\end{equation}

where $\rho$ is the density of the cloud. The black hole mass and the cloud density
have been normalized
to extreme values for a putative torus (Krolik \& Begelman 1989) in order to obtain the greatest
distance. The black hole mass M$_\bullet=10^9$ M$_\odot$ is
obtained from mass measurements of central black holes
in nearby galaxies (Kormendy \& Richstone 1998). The typical average density $<\rho>$
in a standard torus can
be evaluated as the ratio between the column density and the thickness of the torus.
Objects of subsample B have N$_H \sim$ several 10$^{23}$ cm$^{-2}$, then  $\rho_{AV} \sim 10^5$
cm$^{-3}$. Given the clumpiness of the torus, the actual density of a cloud can be higher, however
the value  $\rho = 10^6$  cm$^{-3}$
can be regarded as an upper limit. Even though most of the physical parameters
in the equation above are poorly constrained, several conclusions can nevertheless be drawn.
First, the shorter timescale N$_H$ variations ($\sim$ 2 month or less) must be due to material that
is nearer to the center than about 10$^{17}$ cm, while the radius of the standard model
torus is 1-3 pc (Krolik \& Begelman 1992). The only physical parameter that can reasonably
be larger, assuming different physical conditions, is
the cloud density: for example, photoionization models suggest that 
BELR clouds have $\rho \geq 10^9$ cm$^{-3}$ (Netzer 1990). Even in this case though it is hard to imagine a
reasonable geometry with distances larger than $\sim 10^{17}$~cm, since this would
imply that the absorbing material is confined in a thin slab whose depth is less than
10$^{-4}$ of the distance from the center, but covering  a large solid angle.

These parameters are even better constrained for two of the objects in our sample for which a measurement
of the black hole mass is available: Centaurus A (M$_{\bullet}=2_{-1.4}^{+3}~10^8$ M$_\odot$,
Marconi et al. 2001) and NGC 4258 (M$_{\bullet}=4.2\pm 0.2~10^7$ M$_\odot$, Miyoshi et al. 1995).
Using the fastest N$_H$ variations observed (which are only upper limits to the actual variation timescales)
we obtain:
\begin{equation}
R_{\rm CEN A} \leq 2 \times 10^{15} (\frac{\rho}{10^6 {\rm cm}^{-3}})^2  ~{\rm cm}
\end{equation}
and
\begin{equation}
R_{\rm N 4258} \leq  7\times 10^{16} (\frac{\rho}{10^6 {\rm cm}^{-3}})^2  ~{\rm cm}
\end{equation}

These results pose severe problems to the standard torus model, according to which the 
X-ray absorption is due to cold gas distributed in a toroidal geometry on the parsec
scale.

An alternative possibility, within the standard AGN model (Antonucci 1993)
is that the X-ray absorber
is located in the broad emission line region, much nearer to the central black hole than
the ``standard'' torus.
An indication for an absorber on the BELR scale
is also provided by recent high resolution soft X-ray spectra obtained with XMM-Newton (Sako et al. 2001).
If we assume that the broad line clouds are responsible for the absorption in the X-rays,
we can find a consistent combination of the parameters in Eq. 1. For example, assuming
$\Delta$N$_H \sim 10^{23}$ cm$^{-2}$, M=10$^7$ M$_\odot$, $\rho = 10^9$ cm$^{-3}$,
typical for Seyfert galaxies (Netzer 1990), and t=5 days
we obtain R$\sim 3\times 10^{16}$ cm, $\sim$ 3 light days, a distance
typical for the BELR based on reverberation mapping (Peterson et al. 1997).
Our point can be graphically illustrated as in Fig. 6, where we show that our observed
variability is not compatible with a parsec-scale torus, while a variability of $\sim$ 1 day could
be in agreement with an absorber located in the BELR.

The typical column density of the obscuring clouds also lies in the regime of BELR clouds.
This variability consequently requires that the circumnuclear absorber is not homogeneous,
but clumped into several clouds. Assuming Gaussian fluctuations, the average number of
clouds along the line of sight, N$_C$, should be of the order $(1/f)^2$. The right hand axis of Fig. 1
shows this number to be of order a few for most of the sample. In no case though, even for
N$_C <1$, did N$_H$ drop below 5$\times 10^{21}$ cm$^{-2}$, well above the lowest detectable value for these
observations Therefore, despite the large variability in N$_H$, there are no changes from type 2 to type 1.
From Fig. 1 we estimate that the average column density of the single clouds, N$_{HC}=<N_H>/N_C$,
range from
a few 10$^{21}$ cm$^{-2}$ in low-N$_H$ objects, up to several 10$^{23}$ cm$^{-2}$ in high N$_H$
objects. We report the estimates of N$_{HC}$ in the last column of Table 2. The
N$_{HC}$ distribution is plotted the histogram in Fig. 7.

However, the standard model for AGNs fails to explain X-ray absorption through the broad line
clouds because it would predict several unseen features. Photoionization
models require that in type 1 AGNs the covering factor of the broad line clouds (assuming an
isotropic distribution around the central black hole) is typically 10\% and certainly
much lower than 100\%. We would then expect that (1) 10\% of
type 1 AGNs would be X-ray absorbed, and (2), repeated observations of the same object would
reveal, in 10\% of the cases, an X-ray absorbed spectrum. At present about 20 bright
Seyfert 1s have been extensively studied in the X-rays for several years, by means of
several X-ray observatories (Turner \& Pounds 1989, Nandra \& Pounds 1989, Nandra et al. 1997),
and these effects have been observed only in one case (NGC 3516, Costantini et al. 2000).
In the optical, where many
more objects have been extensively observed, there are only few known cases of transition from type 1 to
type 2 or vice versa (NGC~7603, Tohline \& Osterbrock 1976; MKN~1018, Cohen et al. 1986; NGC~7582,
Aretxaga et al. 1999; for other cases see references in  Aretxaga et al. 1999).
We conclude that the standard AGN model is not able to reproduce the observed
variability of X-ray absorption in Seyfert 2 galaxies.

\subsection{Alternative scenarios}

From the discussion above we conclude that the observed X-ray variability must be
explained by an absorber with the following properties: (1) close to the central engine
(unless moving faster than Keplerian), (2) clumped (to explain the N$_H$ variability,
with N$_C \sim 10$), (3) covers a significant fraction, $\sim 0.8$, of the solid
angle (in order to reproduce
the 4:1 ratio between Seyfert 2s and Seyfert 1s), and (4) since AGN rarely change from type 2 to type 1,
the absorber cannot be spherically symmetric so that the 80\% of lines of sight covered must
always remain the same.

\subsubsection{Type 2 AGNs from an excess of broad line clouds}

A straightforward  extension of the standard unified model that can reproduce the right X-ray
absorption properties is to suppose that in Seyfert 2s
the BELCs are, somewhat paradoxically, {\em much more numerous}
than in Seyfert 1s, so that the covering factor
is 100\% and on average several clouds cover the line of sight.

We can repeat the argument used above to rule out the parsec-scale torus, to constrain the
variability timescales within this model: assuming that the absorber is made by BELCs ($\rho \sim
10^9$ cm$^{-3}$) moving with
Keplerian velocity,
we can obtain a consistent distance from the center only assuming variations
on timescales of days.  Assuming a typical velocity of 5000 km s$^{-1}$ for the BELCs, the crossing
time of a cloud is $t \sim 6 (\frac{\rho}{10^9 {\rm cm}^{-3}})^{-1}(\frac{N_H}{10^{22} {\rm cm}^{-2}})$ days,
and the distance from the center is R$\sim 5\times 10^{16} \frac{M_\bullet}{10^8 M_\odot}$ cm.

Rewriting Equation 3 for a BELR absorber, and using a typical N$_H$ variation for the
high-N$_H$ subsample, we have:

\begin{equation}
R \sim 4\times 10^{16} \frac{M_\bullet}{10^8M_\odot}~(\frac{\rho}{10^9 {\rm cm}^{-3}})^2~(\frac{t}
{{\rm 1~day}})^2~(\frac{N_H}{5 \times 10^{22}{\rm cm}^{-2}})^{-2} ~{\rm cm}
\end{equation}

Such a model is novel because would ascribe the differences between
type 1 and type 2 AGNs to physical properties, rather than to orientation effects.
Orientation is reintroduced because the X-ray absorber cannot be spherically distributed around the center
(because of the shortage of objects changing from type 1 to type 2 and vice versa, as outlined in the previous
Section).

The simplest possible geometry for a non-spherically symmetric absorber is axisymmetric:
a bi-cylinder or a bi-cone.
This geometry is appealing because the same structure has been recently proposed by Elvis (2000)
in order to explain the
absorption and scattering properties of type 1 AGNs. According to this model, some disk instability
generates a two-phase wind from a narrow range of disk radii:
the warm phase, with temperatures of the order of
$\sim 10^5-10^6$ K, is responsible for most of the scattering phenomena observed in AGNs,
and for both the UV narrow absorption lines and the X-ray warm absorbers
seen in a significant fraction of type 1 AGNs by the Hubble
Space telescope (Crenshaw et al. 1999) and ASCA (Reynolds 1998).
The cold phase of the wind is formed by the BELCs (Figure 8).
A simple extension of this model could be
that in type 2 AGNs the BELCs confined the warm wind are far more common and
completely cover all lines of sight, i.e. on average more than one cloud is
present along each line of sight. The fraction of matter
in the different phases of multi-phase media, e.g. the ISM, is not predictable with current theory,
and may be history dependent. Some alteration in the seed medium produced by a change
in the disk instability at least provides a plausible site for such a bi-modal state to originate.
A variation of this is that a larger radial thickness of
the wind might produce the dichotomy between type 1 and type 2 AGN.
In this case, the average distance of the center of the X-ray absorber could
be significantly
larger than the distance of the BELR, since the BELCs will be only those
located close to the inner edge of the wind (Fig. 8). Indeed, indications of
a large thickness of the BELR come also from reverberation mapping studies,
according to which low ionization lines, as MgII $\lambda$2800\AA, are emitted by a
region several times farther from the center than that emitting high ionization
lines like HeII $\lambda$1640\AA~and NV  $\lambda$1240\AA~(Peterson 1993).

An interesting prediction of this model is that
the change from type 2 to type 1 could occasionally happen, when the line of sight
is freed from clouds, because of random motions. In these cases,  broad lines in the
optical and soft X-rays outburst would be observable. The duration of these breaks in
the clouds is of the same order of the crossing time of a cloud along the
line of sight. Assuming the parameters used in
Eq. 2, and a column density for a single cloud of $\sim 10^{22}$ cm$^{-2}$, we can estimate
this time to be t$\sim$ 1 day, too short to be caught with anything but the most intense monitoring.
The probability $P$ that this happens depends on the average
number of clouds N$_C$ along the line of sight. Assuming  N$_C$=4 and N$_C$=10 (two plausible values,
see Fig. 1) we have P$\sim$1\% and P$\sim$ 0.01\%, respectively. A detailed study of this
case will be presented in Nicastro et al. 2001 (in preparation).

A problem with this ``many clouds'' model comes from to the optical emission
of Seyfert~2s. AGNs that are obscured only by the broad line cloud gas should not be
obscured in the optical continuum, but only in the resonant lines themselves. But Seyfert 2s continua
are weak (Koski 1978).
Instead Seyfert 2s clearly suffer extinction due to dust (Peterson 1997). Some dust must then be
present in the BELCs of type 2 AGN. Certainly the outer layers of clouds are heavily shielded from the
ionizing continuum and could be the locus of dust formation, although
the timescales required in our case are probably too short.
Another possibility is that the material inflowing from the host galaxy and feeding the AGN
is dusty, and part of the dust is transferred to the external part of the wind, that is much
thicker and farther from the center than in the ``optically thin'' model of Elvis 2000. Therefore,
part of the dust could survive until it reaches the wind. The amount of inflowing material could be
the physical quantity that tunes the thickness of the wind, thus determining the type 1
or type 2 classification.
This possibility is interesting,  given the growing
evidence suggesting a low dust to gas ratio in the absorbing medium of many
AGNs (Maiolino et al. 2000, Risaliti et al. 2001).
A dust poor absorber, located close to the central black hole, has already been suggested by
Granato et al. (1997). These authors show that the infrared emission of Seyfert 2s is best reproduced
assuming a low dust-to-gas ratio, and suggest that most of the X-ray absorber is located inside
the AGN sublimation radius.

Our ``intrinsic'' model for the difference
between type 1 and type 2 AGN also explains
the otherwise puzzling lack of strong infrared dust emission observed in most Seyfert 1s.
Edelson et al. (1987) show that only about 1/3 of Seyfert 1s show strong dust contribution to
infrared emission in IRAS. This observational
fact is hard to include in the standard unified models: Seyfert 1s should have the
same obscuring tori as Seyfert 2s, and therefore their infrared emission should be of the same
order of that of Seyfert 1s, with respect to the bolometric intensity. The model we proposed
predicts that orientation apply only to objects with ``many'' clouds, that are type 2 if seen through the wind
and type 1 (with high IR dust emission) if seen pole-on. Objects with ``few'' clouds are type 1 along every line
of sight, and are not expected to be strong IR emitters. Finally, this scenario is appealing because
it predicts a link between the absorbing column density and the amount of material inflowing from
the host galaxy, in agreement with the finding that
heavily absorbed AGNs are preferentially hosted in strongly barred galaxies (Maiolino et al. 1999).

Within this model we can estimate the fraction $f$ of objects with ``many'' clouds and the average opening angle,
$\alpha$, of the wind (Fig. 8). Using the ratio 4:1 between type 2s and type 1s (Maiolino et al. 1995) and the
fraction 1/3 for the dust emitting type 1s (Edelson et al. 1987) we obtain $\alpha \sim 25^o$
and $f\sim 85$\%. We note that the opening angle of the wind in the model of Elvis 2000 is higher
($\sim~60^o$). However, the presence of a second absorber (see next Section) can reduce this discrepancy.

\subsubsection{A second, large scale, absorber}

The increase in the N$_H$ structure function for low N$_H$ objects at t$\sim$ 5 years is not explained by
a ``many clouds'', BELR,
origin for the cold X-ray absorber. Instead an additional, second, absorber
is needed, located much farther from the center.
Inhomogeneities in this absorber could
be related to the long term variability on timescales of years revealed in our sample of
objects with N$_H \sim 10^{22}$ cm$^{-2}$.
Assuming a column density variation of 5$\times10^{21}$ cm$^{-2}$ (typical for the long-timescale
variations observed in the low-N$_H$ subsample), we can parametrize the distance from the center of this
second absorber as follows:

\begin{equation}
R \sim 10^{19} \frac{M_\bullet}{10^8M_\odot}~(\frac{\rho}{10^6 {\rm cm}^{-3}})^2~(\frac{t}
{{\rm 5~yr}})^2~(\frac{N_H}{5 \times10^{21}{\rm cm}^{-2}})^{-2} ~{\rm cm}
\end{equation}

This absorber can be compatible with the standard torus model (R$\sim 10^{19}$ cm), in objects with
a lower M$_\bullet$ or with a lower cloud density.

The existence of a second absorber is needed by the BELC-origin model, that, in the original
version of Elvis (2000)  implies a covering
factor of $\sim$ 0.5 and so cannot reproduce the high ratio between Seyfert 2s and Seyfert 1s, which is known
to be $\sim 4$ for nearby objects (Maiolino \& Rieke 1995). Moreover, strong suggestions of the existence of
two distinct absorbers come from the analysis of the X-ray spectra of several nearby AGNs (Malaguti et al. 1999,
Turner et al. 2000, Vignali et al. 1998)
The orientations of the two absorbers are likely unrelated, since neither radio jet axes nor
emission line bi-cones (Pogge 1989) align with galaxy minor axes (Ulvestad \& Wilson 1984).
There will then be objects obscured only by one of the two absorbers. This obviously would
increase the fraction of obscured lines of sight.
Objects obscured only by the more distant medium cannot have column density variations
in timescales of days. This is in agreement with the measurements described in the
previous sections.

A consequence of this second absorber is that the ratio between free and wind-covered lines
of sight can be higher than predicted in the previous Section, since some fraction of the objects are type 2
because they are covered by the farther absorber. This would reduce the opening angle
discrepancy between our
model and that of Elvis (2000).
We can give a quantitative estimate of this effect if we assume
that intermediate type objects ($\sim$ sample A) are those observed through the outer torus, but not through
the wind. We therefore have a new, lower, ratio between type 1.8-2 and type 1-1.5 objects,
and an estimate of the covering factor of the
outer obscurer from the ratio between intermediate and type 1 Seyferts. With these number we
calculate an half-opening angle of 35$^o$ and a fraction of objects with ``many clouds''
$f$=75\%. The half-opening angle is still significantly lower that in the model by Elvis (2000). However,
we note that in the Elvis model the wind is turned by radiation pressure. We expect that the
wind in our model, more massive that that of Elvis (2000), is harder to turn by radiation pressure.

\section{Conclusions}
In this paper we have shown that large amplitude (20-100\%), rapid ($\lesssim$~months)
variability of X-ray absorbing column density is an almost ubiquitous
property of Seyfert 2 galaxies. From our compilation of 25 objects with at least 2 observations in the hard X-rays
(2-10 keV) we found that in 21 cases the N$_H$ varies at a confidence level greater than 90\%.
We demonstrated that the variations of N$_H$ are not related to the underlying X-ray flux and so are not
due to variations in the ionization state of the absorber. Variations
of the amount of absorbing gas along the line of sight is the only obvious alternative. Our result
requires the absorbing gas to
be clumpy (in order to reproduce the observed variability) and its distribution around the
black hole cannot be spherically symmetric (since changes from type 2
to type 1 are not observed in the X-rays).

For a well observed subsample of 11 sources with at least 5 observations
in the hard X-rays, the structure function
shows that the typical variation timescales are shorter than several months. The data do not probe shorter
timescales well, so this is an upper bound.
The shortest measured variation is less than 1 year
for 70\% of the sources for which this can be measured.

From the structure function of the low N$_H (<10^{23}$ cm$^{-2}$) sample we also find
an indication of the presence of a larger scale absorber,
responsible for N$_H$ variations on timescales of the order of $\sim$ 5 years.

The fast variability is incompatible with the standard
parsec-scale torus of unified models. In order to reproduce the observed X-ray properties, the
absorbers have to be clumpy and close to central black hole (distance R$ < 10^{17}$ M$_\bullet$/M$_\odot$ cm).

We propose a model in which a superabundance of broad emission line region clouds
produces the absorption. The bi-cylindrical geometry of Elvis (2000),
in which the absorber is the cold phase of a wind
arising from the accretion disk reintroduces orientation as a factor in whether we see a type 1 or a type 2 AGN.
The model we proposed also predicts that a change from type 2 to type 1 could occasionally be observable
in the X-rays (further details will be in Nicastro et al. 2001, in preparation).

We found significant variability of N$_H$ on all the timescales that we could
investigate. Moreover, the model we propose predicts N$_H$ variability in timescales of days.
Therefore, this work can be usefully complemented by an analysis of the N$_H$ variations
at shorter timescales. Several observations made with ASCA and BeppoSAX have a sufficient statistics
to look for N$_H$ variations of timescales from $\sim 10,000$ seconds to 2 days.
A more detailed study on this issue will be the subject of
a forthcoming paper (Risaliti et al. 2001, in preparation).
\acknowledgments
This work has made use of data
obtained through the Science Data Center of the Italian Space Agency and  the High Energy
Astrophysics Science Archive Center (HEASARC) on-line archive, provided by NASA-Goddard
Space Flight Center. This work was supported in part by NASA grant NAG5-4808.
\email{aastex-help@aas.org}.

\appendix
\section{New hard X-ray measurements}
22 measurements listed in Table 1 (4 with ASCA and 18 with BeppoSAX) are previously unpublished.
These data  permit the inclusion in the sample of two heavily
obscured objects (NGC 4941 and NGC 3081) and are also important for the statistical significance
of the structure function, since almost
all the spectra have an high signal to noise, allowing n$_H$ measurements among the most precise
in all sample.

The data analysis was performed using a standard technique:

- {\em ASCA data}: we only analyzed the two GIS (0.7-10 keV)
observations, extracting a spectrum from a circular region
of radius 4' centered on the source. The background was obtained from a free region in the same field.
The data were rebinned in order to have at least 20 counts per channel.

- {\em BeppoSAX data}:
we analyzed the data from the LECS (0.1-10 keV) , MECS (1.65-10.5 keV)  and PDS (20-200
keV) instruments. We extracted a spectrum from the LECS and MECS data using a region equal to that
used for the ASCA data. We used the PDS spectrum provided by the SAX Science Data Center (SDC).

- {\em Models}: we fitted the data with a baseline model composed by an absorbed power law plus a
thermal component. In many cases, given the high signal to noise of the spectra, several featured
are clearly non fitted. In order to obtain a good fit, we added extra components to the baseline model
(a Gaussian emission line to fit the iron K$\alpha$ emission at E$\sim$6.4 keV, a cold reflection component,
an high energy cutoff for sources observed with BeppoSAX). We give a detailed description of these
fits in another paper (Risaliti  2001, in preparation). The important point  for the measures
used in this work is that in all cases we finally obtained a statistically acceptable fit, and the N$_H$
measurements are not significantly affected by the details of the fit of the other spectral components.

\clearpage

\figcaption[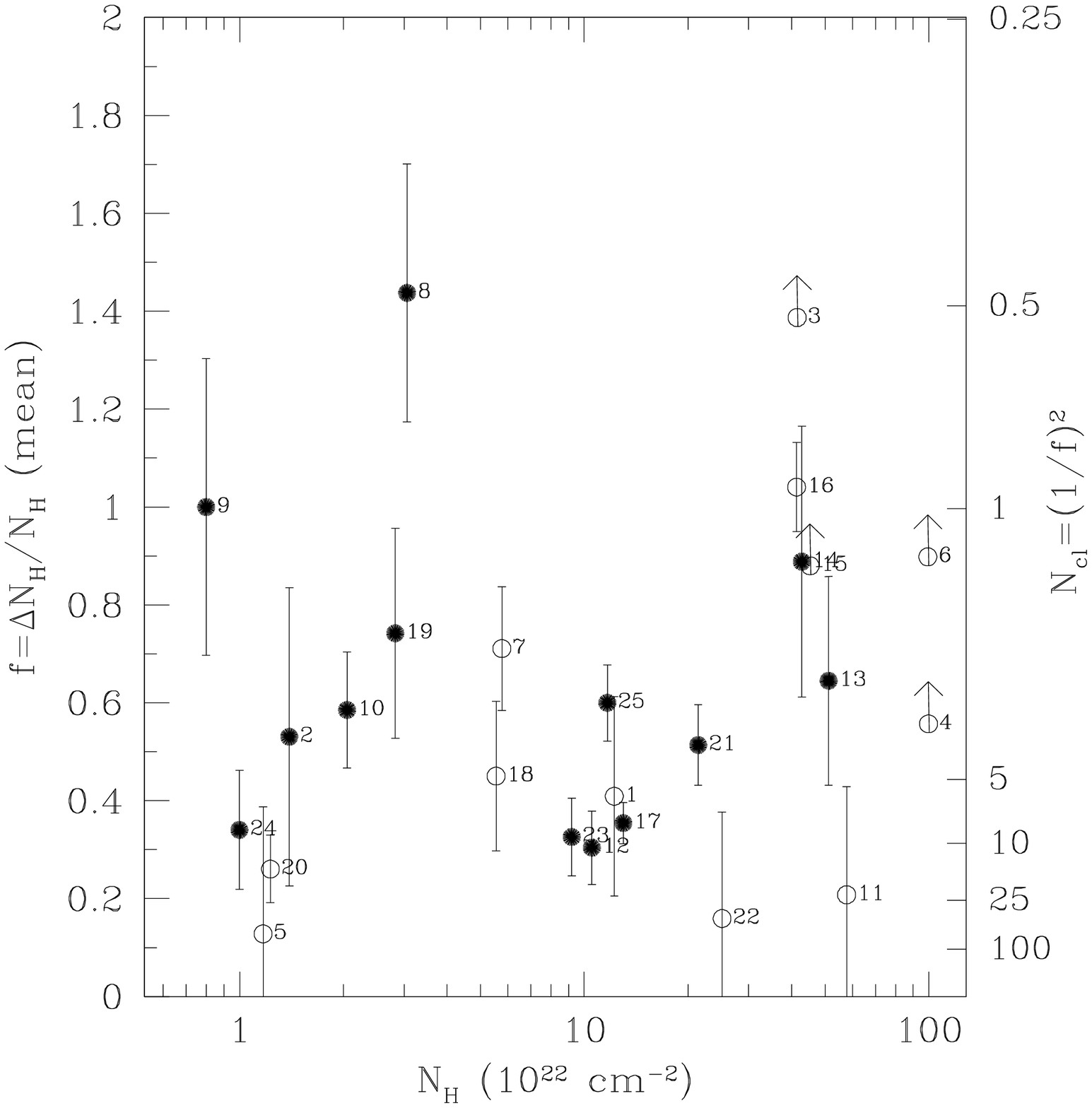]{Ratio between the variation of N$_H$ and the mean N$_H$ for
all the Seyfert 2s with multiple hard X-ray observations. Empty circles are used for sources with
only 2 or 3 observations in the hard X rays, full circles for sources with 4 or more observations.\label{fig1}}
On the right y-axis we report the expected average number of clouds along the line of sight, assuming
Poissonian fluctuations.

\figcaption[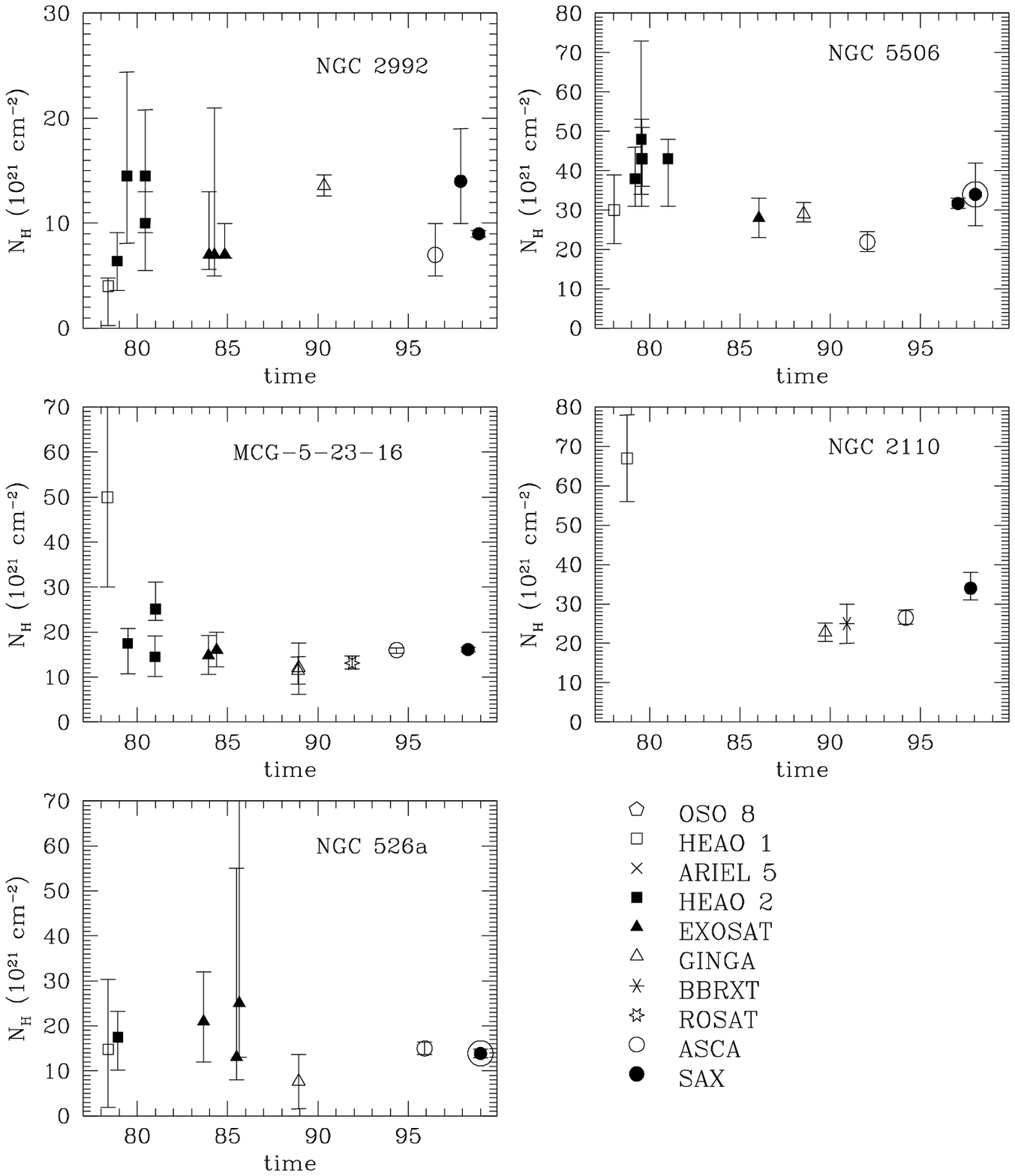]{N$_H$ light curves for a subsample of 5 Seyfert 2s with
N$_H \sim 10^{22}$ cm$^{-2}$ and multiple hard X-ray observations.
\label{fig2}}

\figcaption[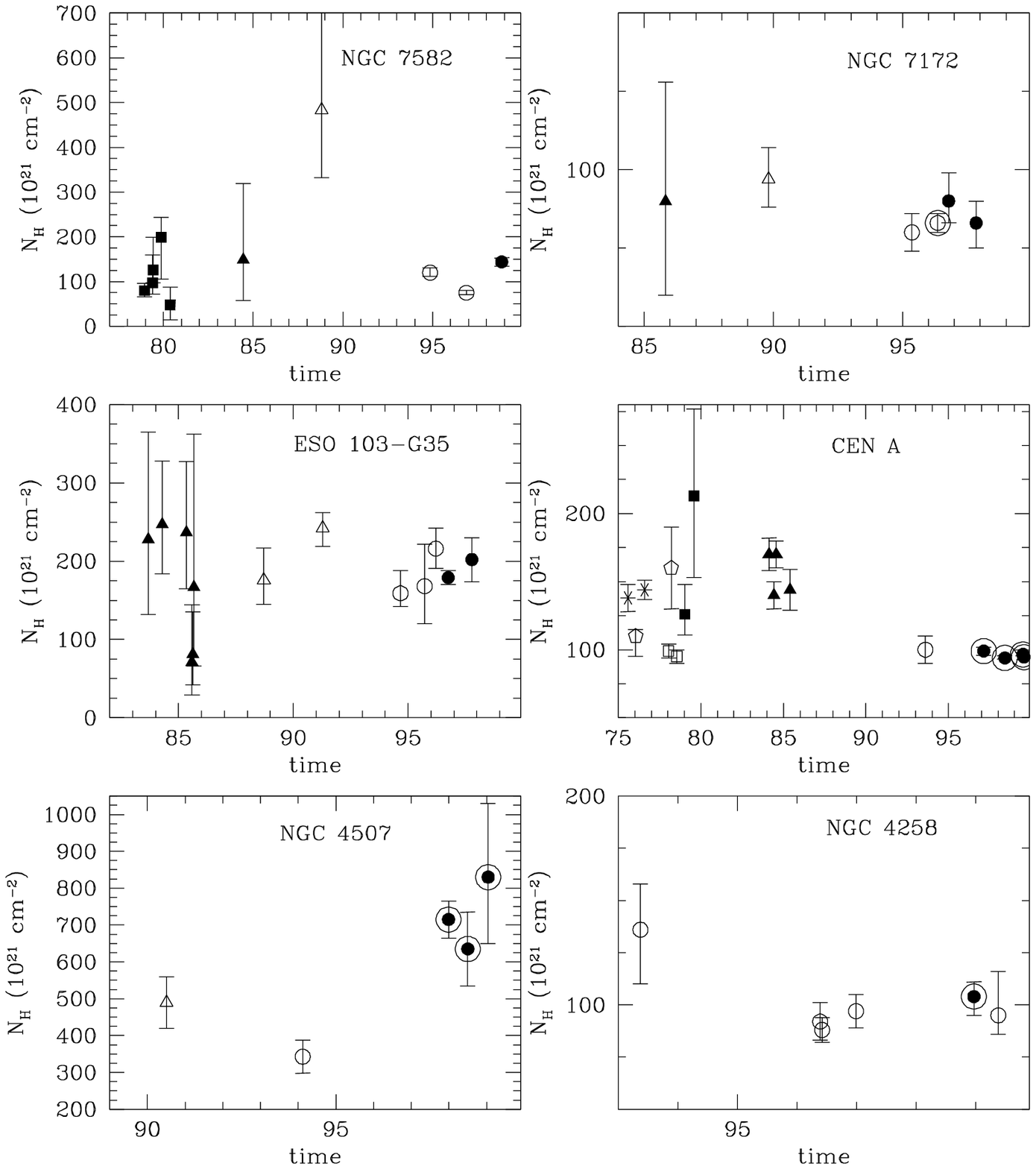]{N$_H$ light curves for a subsample of 5 Seyfert 2s with
N$_H \sim 10^{23}$ cm$^{-2}$ and multiple hard X-ray observations.
\label{fig3}}

\figcaption[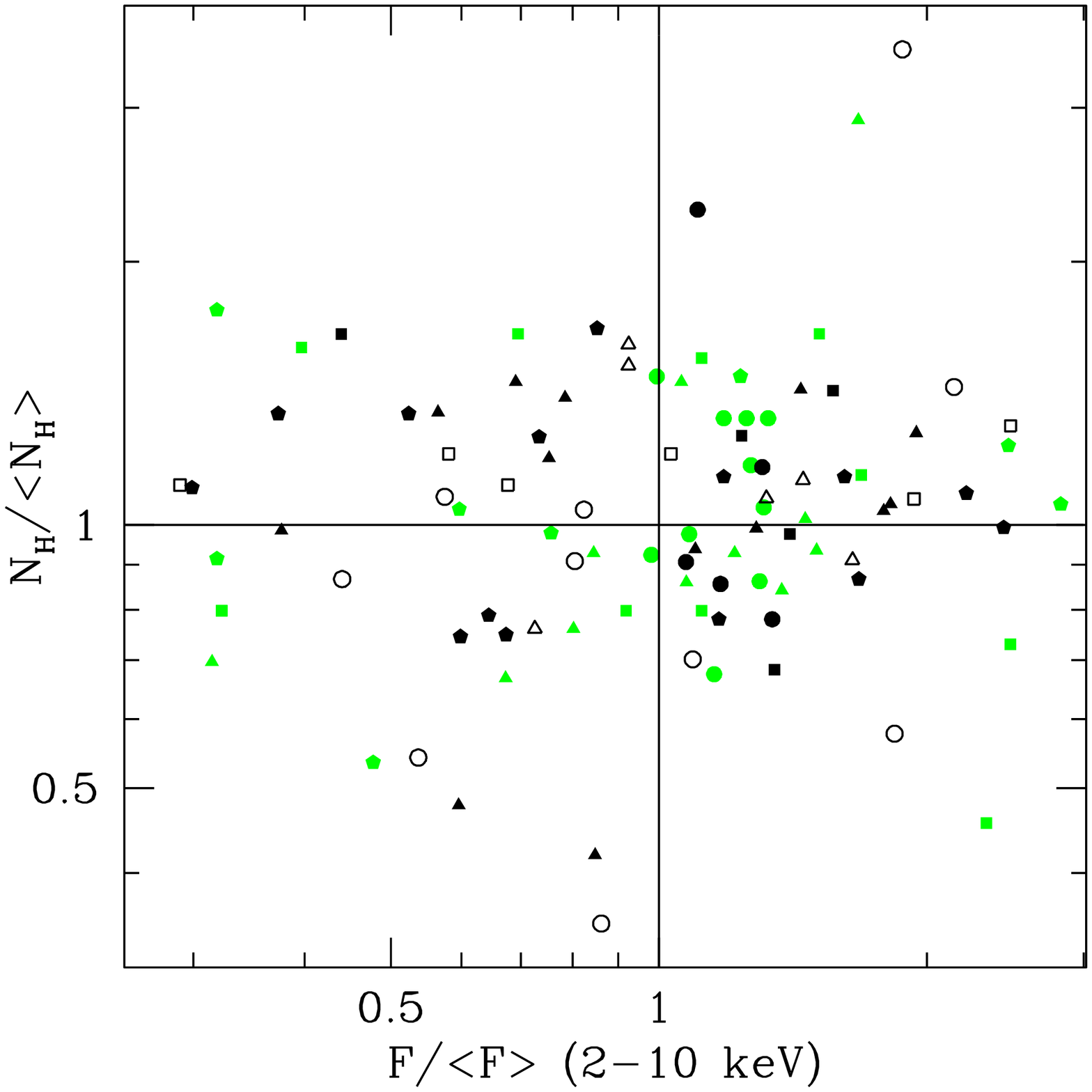]{N$_H$ versus 2-10 keV flux for the observations of the well observed subsample.
Both N$_H$ and flux
are normalized to the average value for each source. Different simbols are used for each of the
11 sources of the subsample.}

\figcaption[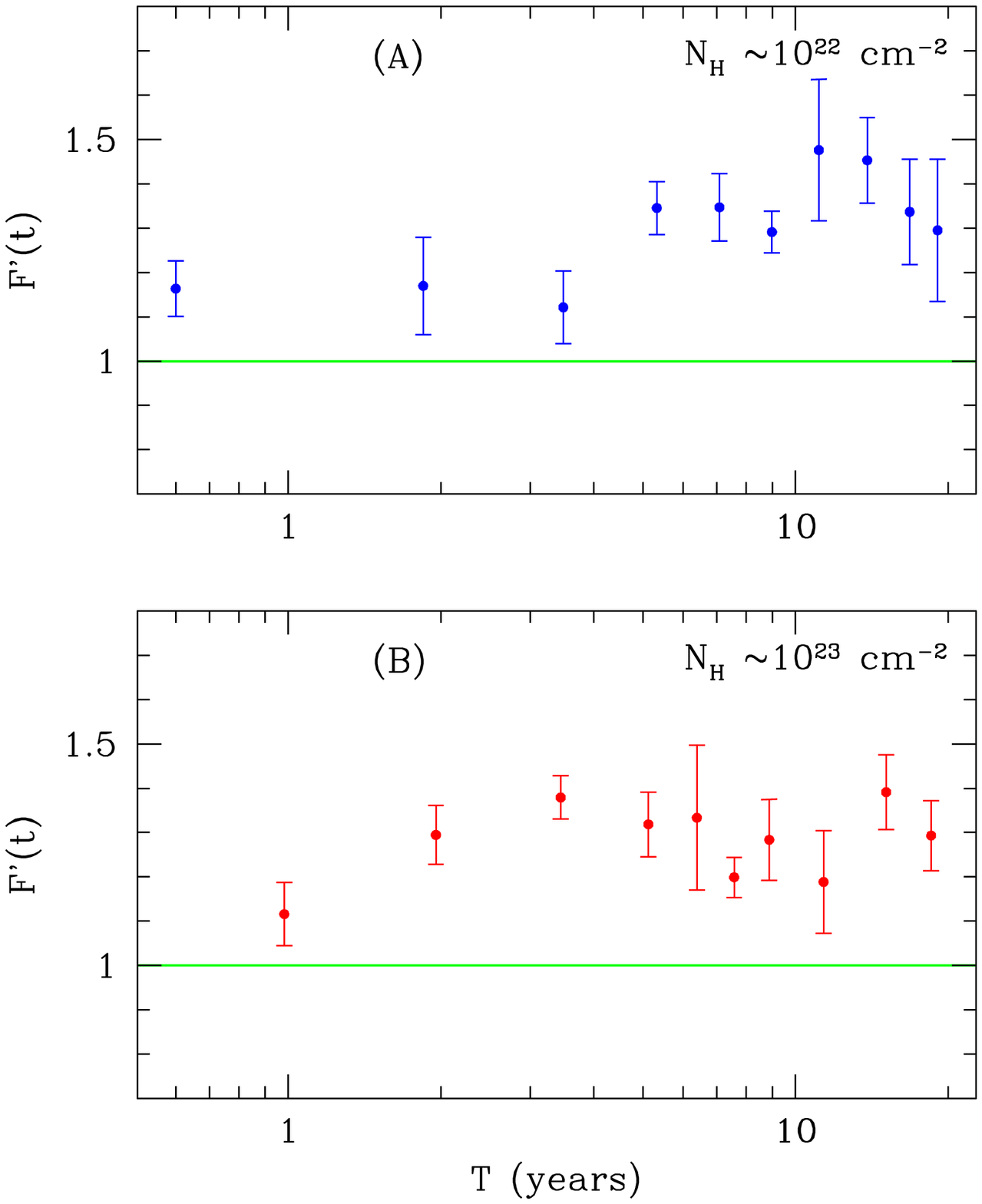]{Structure function for the Column density variations (see text
for definitions). Panel (A): subsample with N$_H \sim 10^{22}$ cm$^{-2}$.
Panel (B): subsample with N$_H \sim 10^{23}$ cm$^{-2}$. \label{fig4}}

\figcaption[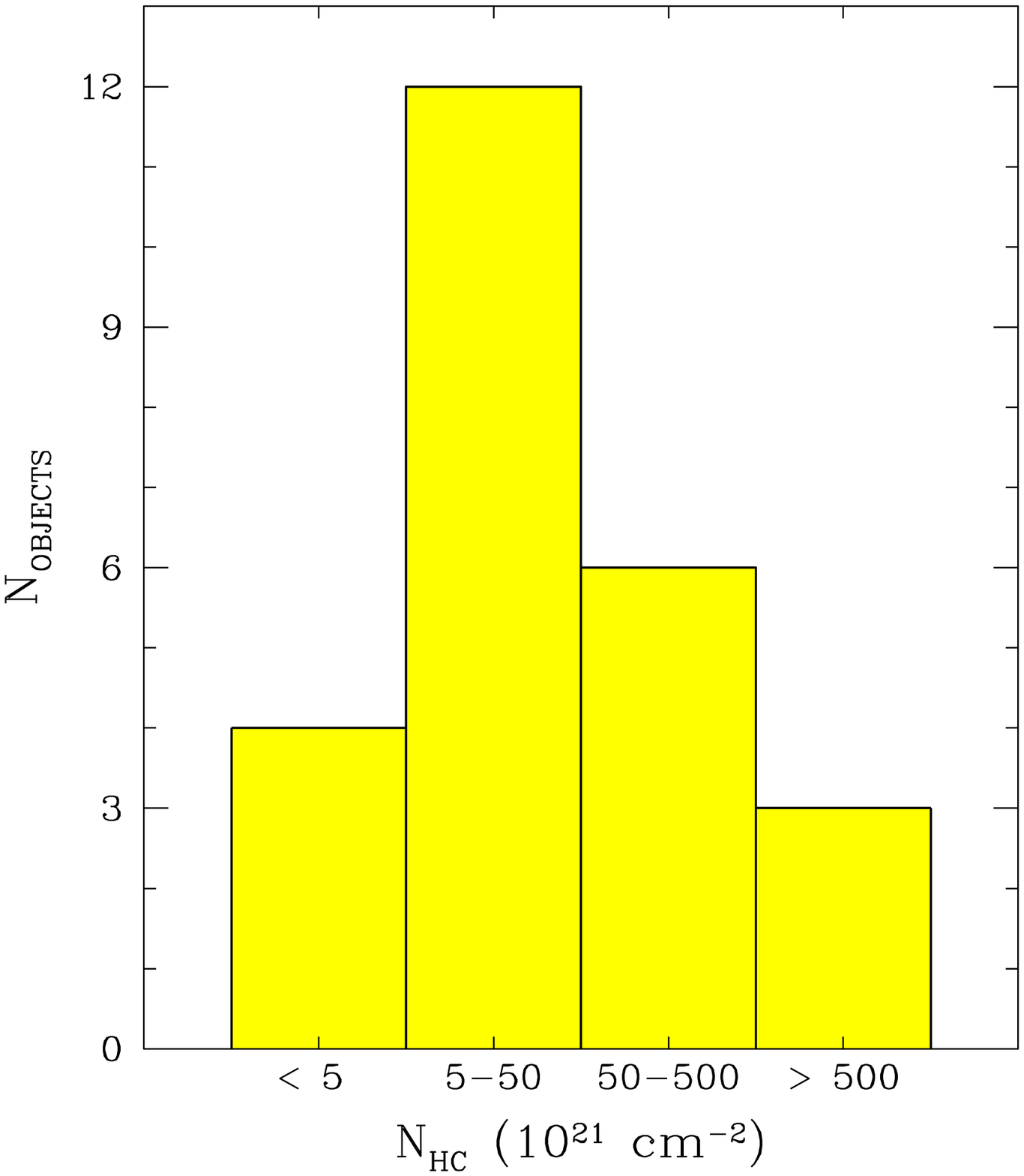]{Distance from the center  (in units of R$_S$) versus cloud density.
The box on the upper-left of the diagram is the region of the parameter space occupied
by a standard parsec-scale torus. The box on the bottom-right represents the BELR.
The three lines are obtained from Equations 3, 6 and 7.
The shortest variations observed, with timescales of a few months, rule out the
parsec-scale torus scenario. Instead, timescales of $\sim$ 1 day are in agreement with
the hypothesis of an absorber located in the BELR.}

\figcaption[f7.eps]{Column density distribution for the single absorbing clouds,
estimated as the ratio of the average N$_H$ and the expected number of clouds, N$_C$,
along the line of sight.}

\figcaption[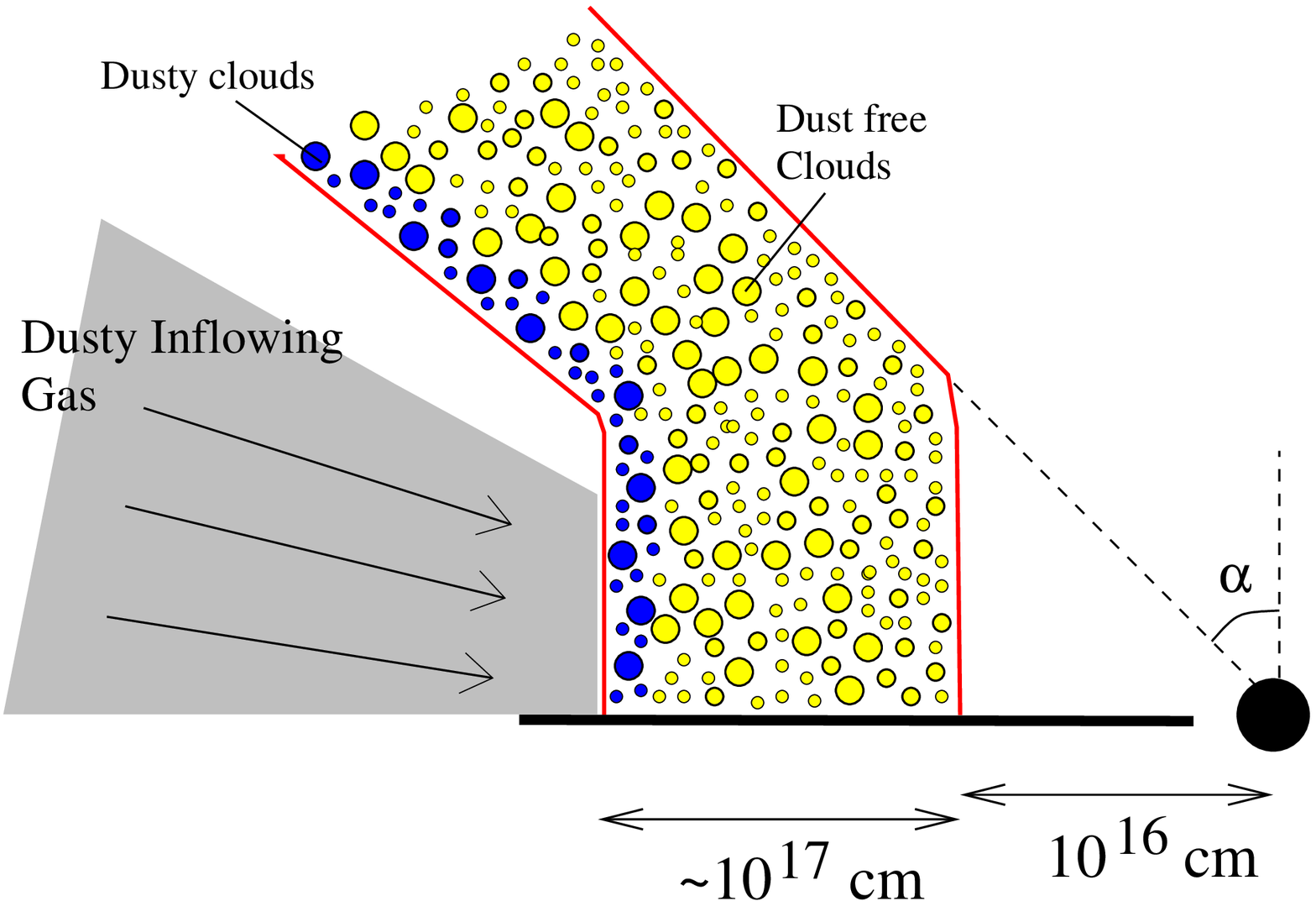]{A simple model, derived from Elvis (2000) to explain Seyfert 2 X-ray
absorption properties. Both broad emission lines (emitted by the inner clouds) and the
X-ray continuum (emitted by the central region of the accretion disk) are absorbed by
the clouds inside the wind.
The column density variability timescale is the average crossing time
of a cloud along the line of sight. Note that the dimensions are different from the
original Elvis 2000 model, where the radial size of the wind is $\sim 10^{16}$ cm (see
text for details).}

\begin{figure}
\plotone{f1.eps}
\end{figure}

\begin{figure}
\plotone{f2.eps}
\end{figure}

\begin{figure}
\plotone{f3.eps}
\end{figure}

\begin{figure}
\plotone{f4.eps}
\end{figure}

\begin{figure}
\plotone{f5.eps}
\end{figure}

\begin{figure}
\plotone{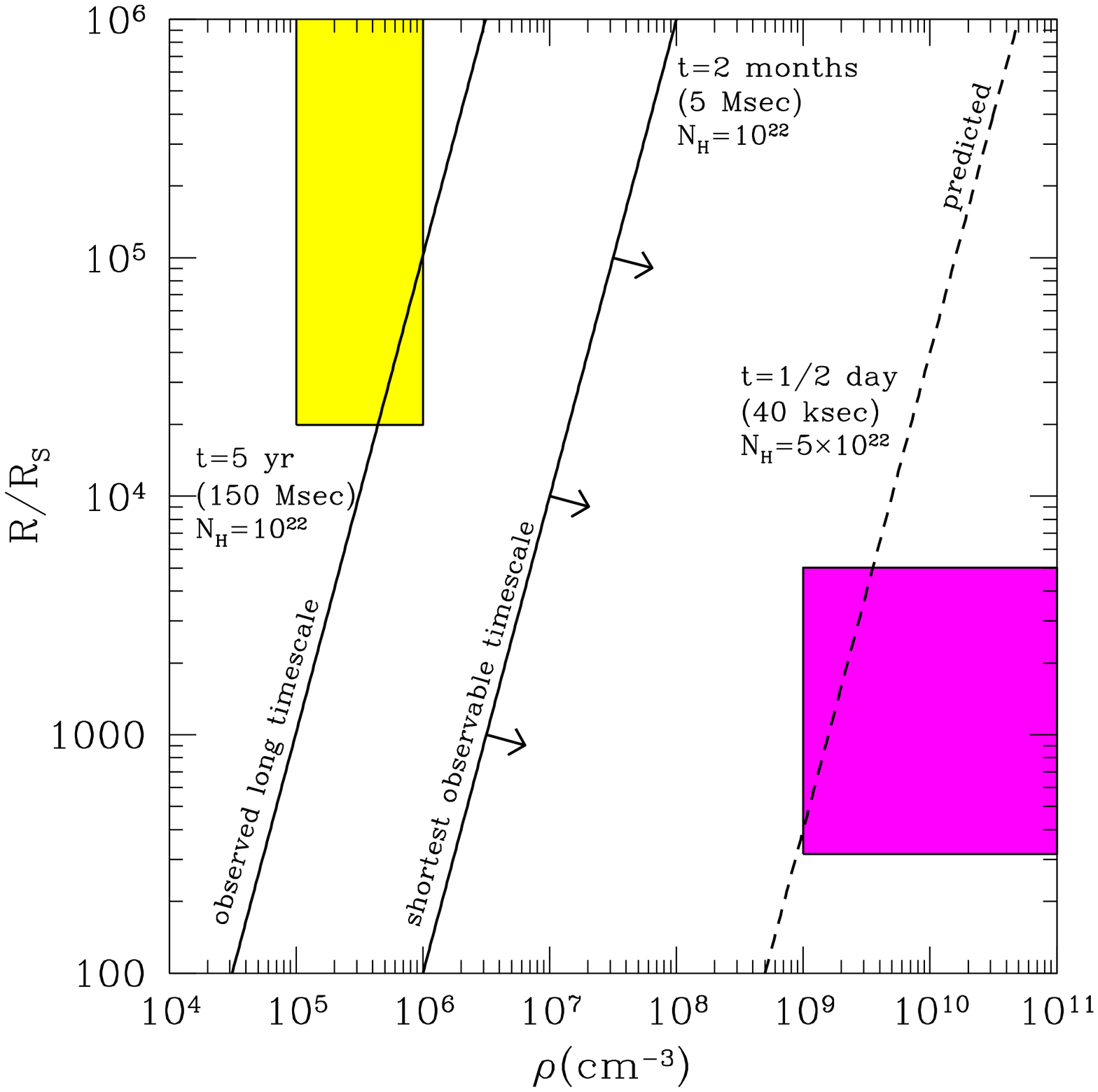}
\end{figure}

\begin{figure}
\plotone{f7.eps}
\end{figure}

\begin{figure}
\plotone{f8.eps}
\end{figure}

\begin{deluxetable}{lrrrrrc}
\tablewidth{0pt}
\tablecaption{Data for our sample of Seyfert 2s}
\tablehead{
\colhead{\#} &
\colhead{Name} & \colhead{Obs. date}      &
\colhead{Instrument} & \colhead{N$_H$\tablenotemark{a}} & \colhead{F(2-10 keV)\tablenotemark{b}}
& \colhead{Ref.}}
\startdata

1 &MKN 348          & 13.06.1987 & GINGA 	& 106$^{+31}_{-26}$  	& 2.2	& 2 \\
& Sy 2		 & 04.08.1995 & ASCA 	& 160$^{+20}_{-10}$	& 0.8	& 24 \\
2 & NGC 526a$^{(A)}$ & 23.05.1978 & HEAO 1 & 15$^{+15}_{-13}$  	& 1.5	& 6 \\
& Sy 2/1.5	 & 09.12.1978 & HEAO 2 	& 17.5$^{+6}_{-7}$   	& 6.2	& 9 \\
&		 & 05.08.1983 & EXOSAT	& 21$^{+11}_{-9}$   	& 3.1	& 3 \\
&		 & 28.06.1985 & EXOSAT	& 13$^{+42}_{-5}$   	& 0.8	& 3 \\
&		 & 23.08.1985 & EXOSAT	& 25$^{+53}_{-12}$ 	& 0.8	& 3 \\
&		 & 15.12.1988 & GINGA 	& 7.6$^{+6}_{-6}$		& 1.2	& 2 \\
&		 & 30.11.1995 & ASCA 	& 15$^{+1.4}_{-1.4}$	& 7.1	& 4 \\
&		 & 01.01.1999 & SAX 	& 16$^{+2}_{-2}$		& 1.9	& 1 \\
3 & NGC 1365      & 25.01.1995 & ASCA 	& $>10000$	   	& --	& 12 \\
& Sy 1.8	 & 12.08.1997 & SAX 		& 400$^{+40}_{-50}$ 	& 2.1	& 13 \\
4 & NGC 1386      & 26.01.1995 & ASCA 	& 280$^{+270}_{-260}$ 	& 0.065	& 12 \\
& Sy 2		 & 10.12.1996 & SAX 	& $>1000$		& --	& 22 \\
5 & IRAS 04575-7537  & 04.10.1990 & GINGA & 12.3$^{+7}_{-4}$  	& 1.6	& 2 \\
& Sy 2		 & 04.11.1996 & ASCA 	& 10.5$^{+1}_{-1}$	& 2.0	& 23 \\
6  & NGC 1808     & 15.10.1990 & GINGA 	& 105$^{+1.4}_{-1.4}$ 	& 0.3 	& 25 \\
& Sy 2		 & 26.02.1994 & ASCA 	& $<2.5$		& 0.09	& 4 \\
7 & IRAS 05189-2524  & 15.02.1995 & ASCA & 44$^{+4}_{-4}$	& 0.6	& 10 \\
& Sy 2		 & 03.10.1999 & SAX 	&  85$^{+8}_{-8}$		& 0.7	& 10 \\
8 & NGC 2110$^{(A)}$ & 09.10.1978 & HEAO1 & 67$^{+11}_{-11}$   	& 3.3	& 6 \\
& Sy 2   	 	 & 26.09.1989 & GINGA 	& 22.8$^{+2}_{-2}$	& 4.0	& 2 \\
&		 & 06.12.1990 & BBRXT 	& 25$^{+5}_{-5}$		& 3.5	& 7 \\
&		 & 12.03.1990 & ASCA 	& 26.5$^{+2}_{-1.6}$	& 3.2	& 4 \\
&		 & 14.10.1997 & SAX 	& 34$^{+3}_{-3}$ 		& 3.9	& 8 \\
9 & NGC 2992$^{(A)}$ & 24.05.1978 & HEAO 1 & 4.0$^{+0.8}_{-4}$  	& 9.4	& 32 \\
& Sy 1.9		 & 22.11.1978 & HEAO 2	& 6.4$^{+2.7}_{-2.8}$ 	& 10.0  & 9 \\
&		 & 02.06.1979 & HEAO 2 	& 14.5$^{+10}_{-6}$ 	& 2.8	& 9 \\
&		 & 06.06.1980 & HEAO 2	& 10$^{+3}_{-4.5}$  	& 6.8	& 9 \\
&	    	 & 09.06.1980 & HEAO 2	& 14.5$^{+6.3}_{-5.4}$	& 6.1	& 9 \\
&		 & 18.12.1983 & EXOSAT	& 7$^{+6}_{-1.4}$		& 1.3	& 9 \\
&		 & 06.04.1984 & EXOSAT	& 7.8$^{+14}_{-1.9}$	& 4.5	& 9 \\
&		 & 09.11.1984 & EXOSAT	& 7$^{+3}_{-0.3}$		& 3.7	& 3 \\
&		 & 30.04.1990 & GINGA 	& 16.1$^{+4}_{-7}$	& 1.57 	& 2 \\
&		 & 30.06.1996 & ASCA 	& 7$^{+3}_{-2}$ 		& 0.35	& 4 \\
&		 & 30.11.1997 & SAX 	& 14$^{+5}_{-4}$		& 0.7	& 14 \\
&		 & 30.11.1998 & SAX 	& 9$^{+0.3}_{-0.3}$	& 7.8	& 14 \\
10 & MCG-5-23-16$^{(A)}$ & 12.05.1978 & HEAO 1 & 50$^{+20}_{-20}$ 	& 11.7	& 6 \\
& Sy 2		 & 23.06.1979 & HEAO 2 	& 18$^{+3}_{-7}$		& 10.2	& 9 \\
&		 & 31.12.1980 & HEAO 2 	& 14.5$^{+5}_{-4}$   	& 9.6	& 9 \\
&		 & 04.01.1981 & HEAO 2 	& 25$^{+6}_{-3}$		& 7.4	& 9 \\
&		 & 13.12.1983 & EXOSAT	& 15$^{+4}_{-4}$		& 7.5	& 3 \\
&		 & 23.04.1984 & EXOSAT	& 16$^{+4}_{-4}$		& 5.9	& 3 \\
&		 & 30.11.1988 & GINGA 	& 19.4$^{+4}_{-5}$	& 4.7 	& 2 \\
&		 & 05.12.1988 & GINGA 	& 18.1$^{+4}_{-8}$ 	& 2.2	& 2 \\
&		 & 27.11.1991 & ROSAT 	& 13.1$^{+1.6}_{-1.3}$	& 5.6	& 15 \\
&		 & 11.05.1994 & ASCA 	& 16$^{+0.5}_{-0.7}$	& 8.5	& 4 \\
&		 & 25.04.1998 & SAX 	& 15.9$^{+0.4}_{-0.4}$ 	& 10.5	& 1 \\
11 & NGC 3081         & 13.05.1996 & ASCA 	& 500$^{+140}_{-120}$	& 2.6	& 1 \\
& Sy 2		 & 20.12.1996 & SAX 	& 640$^{+200}_{-120}$  	& 0.8	& 22 \\
12 & NGC 4258$^{(B)}$ & 15.05.1993 & ASCA & 136$^{+22}_{-26}$	& 1.4	& 34 \\
& Sy 1.9		 & 23.05.1996 & ASCA 	& 92$^{+9}_{-9}$  	& 2.0	& 34 \\
&		 & 06.06.1996 & ASCA	& 88$^{+7}_{-6}$	& 2.2	& 34 \\
&		 & 28.12.1996 & ASCA	& 97$^{+8}_{-8}$	& 2.5	& 34 \\
&		 & 19.12.1998 & SAX 	& 121$^{+7}_{-6}$ 	& 1.4	& 1 \\
&		 & 17.05.1999 & ASCA 	& 95$^{+21}_{-9}$	& 1.1	& 34 \\
13 & NGC 4388	 & 04.07.1993 & ASCA 	& 315$^{+110}_{-100}$	& 3.2	& 20 \\
& Sy 2		 & 21.06.1995 & ASCA 	& 334$^{+100}_{-90}$	& 1.5	& 20 \\
&		 & 09.01.1999 & SAX 	& 380$^{+20}_{-40}$	& 8.2	& 1 \\
&		 & 03.01.2000 & SAX	& 480$^{+180}_{-80}$	& 2.8 	& 1 \\	
14 & NGC 4507$^{(B)}$ & 07.07.1990 & GINGA & 490$^{+70}_{-70}$  	& 5.1	& 2 \\
& Sy 2		 & 12.02.1994 & ASCA 	& 343$^{+45}_{-45}$   	& 4.9	& 4 \\
&		 & 09.04.1997 & SAX 	& 590$^{+80}_{-120}$	& 5.7	& 1 \\
&		 & 19.02.1998 & SAX 	& 540$^{+90}_{-70}$  	& 4.5	& 1 \\
&		 & 04.01.1999 & SAX 	& 710$^{+20}_{-160}$  	& 1.6	& 1 \\
15 & NGC 4941    & 19.07.1996 & ASCA 	& $>1000$ 		& --	& 1 \\
& Sy 2		 & 22.01.1997 & SAX 	& 450$^{+250}_{-140}$  	& 0.3	& 22 \\	
16 & IRAS 13197-164 & 18.07.1995 & ASCA & 760$^{+130}_{-120}$	& 0.6	& 11 \\
& Sy 1.8		 & 22.07.1998 & SAX 	& 330 $^{+50}_{-40}$	& 1.9	& 1 \\
17 & CENTAURUSA$^{(B)}$ & 27.07.1975 & OSO 8 & 139 $^{+9}_{-9}$ 	& 148	& 27 \\
& Sy 2		 & 19.01.1976 & ARIEL 5 & 110$^{+5}_{-15}$  	& 112	& 28 \\
&		 & 01.08.1976 & OSO 8 	& 144$^{+7}_{-7}$   	& 79	& 27 \\
&		 & 15.01.1978 & HEAO 1 	& 99$^{+5}_{-5}$		& 78	& 29 \\
&		 & 15.03.1978 & ARIEL 5 & 160$^{+30}_{-30}$ 	& 49	& 28 \\
&		 & 15.07.1978 & HEAO 1 	& 95$^{+5}_{-5}$	              & 45	& 29 \\
&		 & 15.01.1979 & HEAO 2 	& 126$^{+22}_{-15}$ 	& 163	& 9 \\
&		 & 03.08.1979 & HEAO 2 	& 213$^{+64}_{-60}$ 	& 57	& 9 \\
&		 & 13.02.1984 & EXOSAT	& 164$^{+8}_{-8}$ 	& 35	& 30 \\
&		 & 08.06.1984 & EXOSAT	& 140$^{+10}_{-10}$ 	& 20	& 30 \\
&		 & 30.07.1984 & EXOSAT	& 170$^{+10}_{-10}$ 	& 25	& 30\\
&		 & 29.05.1985 & EXOSAT	& 144 $^{+1.5}_{-1.5}$ 	& 108	& 30 \\
&		 & 14.08.1993 & ASCA 	& 100$^{+10}_{-10}$ 	& 43	& 31 \\
&		 & 20.02.1997 & SAX 	&  100$^{+3}_{-2}$	 & 34	& 1 \\
&		 & 06.01.1998 & SAX 	& 93.0$^{+1.3}_{-1.7}$	 & 43	& 1 \\
&		 & 10.07.1999 & SAX 	& 94.3$^{+2.3}_{-0.8}$ 	 & 40	& 1 \\
&		 & 02.08.1999 & SAX 	& 93.6$^{+1.7}_{-1.8}$ 	 & 40	& 1 \\
&		 & 08.01.2000 & SAX 	& 94.3$^{+1.4}_{-1.1}$ 	 & 39	& 1 \\
18 & NGC 5252    & 28.01.1994 & ASCA 	& 43$^{+7}_{-6}$ 		& 0.75	& 4 \\
& Sy 1.9		 & 20.01.1998 & SAX 	& 68$^{+16}_{-7}$	& 3.8	& 1 \\
19 & NGC 5506$^{(A)}$ & 20.01.1978 & HEAO 1 & 41$^{+22}_{-18}$  	& 6.8	& 6 \\
& Sy 1.9		 & 13.07.1979 & HEAO 2 	& 38$^{+7}_{-7}$		& 8.8	& 9 \\
&		 & 21.07.1979 & HEAO 2 	& 48$^{+25}_{-14}$ 	& 6.9	& 9 \\
&		 & 30.07.1979 & HEAO 2 	& 43$^{+10}_{-12}$  	& 8.7	& 9 \\
&		 & 09.08.1979 & HEAO 2 	& 43$^{+8}_{-7}$      	& 9.2	& 9 \\
 &		 & 06.01.1981 & HEAO 2 	& 43$^{+5}_{-12}$	& 8.2	& 9 \\
&		 & 24.01.1986 & EXOSAT	& 28$^{+5}_{-5}$		& 9.0	& 3 \\
&		 & 19.07.1988 & GINGA 	& 29$^{+3}_{-2}$		& 1.0	& 2 \\
&		 & 06.01.1992 & ROSAT 	& 21.9$^{+2.6}_{-2.4}$	& 8.0	& 15 \\
&		 & 30.01.1997 & SAX 	& 37$^{+2}_{-1}$		& 7.5	& 1 \\
&		 & 14.01.1998 & SAX 	& 39$^{+1}_{-2}$  	& 9.1	& 1 \\
20 & IRAS 18325-5926  & 13.05.1989 & GINGA & 16.3$^{+7}_{-2.6}$  	& 12.8	& 2 \\
& Sy 1.8		 & 11.09.1993 & ASCA 	& 13.2$^{+1}_{-1}$ 	& 1	& 26 \\
&		 & 27.03.1997 & ASCA 	& 10.4$^{+0.4}_{-0.4}$ 	& 2.5	& 1 \\
21 & ESO 103-G35$^{(B)}$  & 04.09.1983 & EXOSAT& 228$^{+137}_{-98}$ & 1.8	& 18 \\
& Sy 2		 & 19.04.1984 & EXOSAT	& 247$^{+81}_{-63}$ 	& 2.2	& 18 \\
&		 & 04.05.1985 & EXOSAT	& 237$^{+90}_{-72}$ 	& 2.5	& 18 \\
&		 & 02.08.1985 & EXOSAT	& 71$^{+73}_{-42}$   	& 2.7	& 18 \\
&		 & 13.08.1985 & EXOSAT	& 81$^{+54}_{-39}$   	& 1.9	& 18 \\
&		 & 04.09.1985 & EXOSAT	& 167$^{+195}_{-101}$ 	& 1.2	& 18 \\
&		 & 23.09.1988 & GINGA   & 176$^{+41}_{31}$	& 5.7	& 19 \\
&		 & 12.04.1991 & GINGA 	& 292$^{+25}_{-24}$	& 4.6	& 2 \\
&		 & 03.09.1994 & ASCA	& 216$^{+40}_{-35}$     & 3.5	& 20 \\
&		 & 26.09.1995 & ASCA 	& 168$^{+54}_{-48}$	& 4.1	& 20 \\
&		 & 18.03.1996 & ASCA 	& 216$^{+26}_{-25}$	& 6.2	& 20 \\
&		 & 03.10.1996 & SAX	& 179$^{+9}_{-9}$	& 5.8	& 21 \\
&		 & 14.10.1997 & SAX 	& 202$^{+28}_{-28}$  	& 2.4	& 21 \\
22 & IC 5063          & 02.10.1990  & GINGA	& 252$^{+95}_{-78}$ 	& 2.3	& 2 \\
& Sy 2		 & 25.04.1994 & ASCA 	& 256$^{+89}_{-70}$	& 2.6	& 4\\
&		 & 27.04.1994 & ASCA 	& 218$^{+22}_{-21}$	& 2.7	& 4 \\
23 & NGC 7172$^{(B)}$ & 26.10.1989 & GINGA & 112$^{+7}_{-8}$ 	& 7.7	& 2 \\
& Sy 2		 & 28.10.1985 & EXOSAT	& 90$^{+38}_{-30}$ 	& 3.2	& 3 \\
&		 & 11.05.1995 & ASCA 	& 82$^{+4}_{-4}$		& 6.0	& 4 \\
&		 & 17.05.1996 & ASCA 	& 77$^{+8}_{-7}$ 		& 2.1	& 1 \\
&		 & 14.10.1996 & SAX 	& 90$^{+9}_{-7}$ 		& 1.8	& 5 \\
&		 & 06.11.1997 & SAX 	& 83$^{+5}_{-5}$ 		& 0.9	& 5 \\
24 & NGC 7314	 & 05.09.1983 & EXOSAT	& 13$^{+35}_{-4}$   	& 1.3	& 3 \\
& Sy 1.9		 & 25.03.1984 & EXOSAT	& 8.2$^{+8.9}_{-0.9}$ 	& 3.2	& 3 \\
&		 & 20.11.1994 & ASCA 	& 8.9$^{+0.4}_{-1.0}$ 	& 3.6	& 4 \\
&		 & 08.06.1999 & SAX 	& 12.2$^{+0.2}_{-1.4}$	& 2.6	& 1 \\
25 & NGC 7582$^{(B)}$ & 14.12.1978 & HEAO 2 & 80$^{+16}_{-14}$  	& 9.6	& 9 \\
& Sy 2		 & 26.05.1979 & HEAO 2 	& 97$^{+62}_{-25}$  	& 5.7	& 9 \\
&		 & 02.06.1979 & HEAO 2 	& 126$^{+73}_{-29}$	& 4.2	& 9 \\
&		 & 21.11.1979 & HEAO 2 	& 199$^{+45}_{-93}$	& 11.2	& 9 \\
&		 & 15.05.1980 & HEAO 2 	& 48$^{+39}_{-34}$  	& 4.5	& 9 \\
&		 & 09.06.1984 & EXOSAT	& 149$^{+170}_{-91}$	& 3.0	& 16 \\
&		 & 25.10.1988 & GINGA 	& 484$^{+254}_{-152}$	& 9.8	& 19 \\
&		 & 14.11.1994 & ASCA 	& 120$^{+11}_{-8}$ 	& 2.3	& 33 \\
&		 & 21.11.1996 & ASCA 	& 75$^{+5}_{-4}$		& 2.8	& 33 \\
&		 & 10.11.1998 & SAX 	& 144$^{+9}_{-10}$	& 4.3	& 17 \\
\enddata
\tablenotetext{a}{X-ray absorbing column density, in units of 10$^{21}$ cm$^{-2}$}
\tablenotetext{b}{2-10 keV intrinsic flux, in units of 10$^{-11}$ erg s$^{-1}$ cm$^{-2}$}
\tablerefs{(1) This work (2) Smith \& Done 1996; (3) Turner \& Pounds 1989; (4) Turner et al. 1997;
(5) Akylas et al. 2001; (6) Weaver et al. 1995; (7) Weaver 1993; (8) Malaguti et al. 1999; (9) Halpern
1981; (10) Severgnini et al. 2000; (11) Bassani et al. 1999; (12) Iyomoto 1997; (13) Risaliti et al. 2000;
(14) Gilli et al. 2000; (15) Mulchaey et al. 1993; (16) Malaguti et al. 1994; (17) Turner et al. 2000;
(18) Warwick et al. 1988; (19)  Warwick et al. 1993; (20) Forster et al. 1999; (21) Wilkes et al. 2000;
(22) Maiolino et al. 1998; (23) Vignali et al. 1998; (24) Awaki et al. 2000; (25) Awaki \& Koyama 1993
(26) Iwasawa et al. 1996; (27) Mushotzky et al. 1978; (28) Stark (1979); (29) Baity et al. 1981;
(30) Morini et al. 1989; (31) Sambruna et al. 1999; (32) Singh et al. 1985;
(33) Xue et al. 1999.; (34) Reynolds et al. 2000}
\end{deluxetable}

\begin{deluxetable}{ccccc}
\tablewidth{0pt}
\tablecaption{Fastest N$_H$ variations.}
\tablehead{
\colhead{Name}           & \colhead{Time interval\tablenotemark{a}}      &
\colhead{variation (\%)\tablenotemark{b} }  &  \colhead{\# Obs.} & \colhead{N$_{HC}$\tablenotemark{c}}
 }
\startdata
CENTAURUS A & 2 months & 46\% & 18 & 2.3 \\
NGC 4258 &  5 months & 10\% & 6 & 1.6 \\
NGC 2110 & 6 months & 15\% & 5 & 2.2\\
NGC 4941 & 6 months & $>$ 81\% & 2 & 27.5 \\
NGC 7582 & 6 months & 122\% & 10 & 3.5 \\
NGC 2992 & 6.5 months & 40\% & 12 & 0.4 \\
ESO 103-G35 & 6.5 months & 19\% & 13 & 5.5  \\
NGC 4507 & 11 months & 25\% & 5 & 19 \\
NGC 4388 & 1 year & 50\% & 4 & 16.5 \\
MCG-5-23-16 & 1.1 years & 94\% & 11 & 0.6 \\
NGC 1386 & 2 years & $>$112\% & 2 & 36 \\
NGC 1365 & 2.5 years & $>$ 85\% & 2 & 30 \\
IRAS 13197-164 & 3 years & 44\%  & 2 & 22\\
NGC 1808 & 3.3 years & 200\% & 2 & 45 \\
NGC 5506  & 3.5 years & 28\% & 11 & 1.1\\
IRAS 18325-5926 & 3.5 years & 24\% & 3 & 0.2\\
NGC 7172 & 3.5 years & 31\% & 6 & 1.5 \\
NGC 5252 & 4 years & 36\% & 2 & 1.3 \\
NGC 7314 & 4.5 years & 31\% & 4 & 0.2\\
IRAS 05189-2524 & 4.6 years & 64\% & 2 & 2.0\\
NGC 256a & 5 years & 94\% & 8 & 0.4\\
MKN 348 & 8 years & 20\% & 2 & 2.5\\
IRAS 04575-7537 & -- & -- & 2 & --\\
NGC 3081 &  -- & -- & 2 & -- \\
IC 5063 & -- & -- & 3 & -- \\
\enddata
\tablenotetext{a}{Time interval between the two closest variations of N$_H$ statistically significant at $>90$\%.}
\tablenotetext{b}{Variation with respect to the average N$_H$, calculated using the best fit values in Table 1.}
\tablenotetext{c}{Estimated N$_H$ of a single cloud, in units of 10$^{22}$ cm$^{-2}$ (see text for details).}
\end{deluxetable}

\end{document}